\documentstyle[11pt,newpasp,twoside,epsf]{article}
\def\simless{\mathbin{\lower 3pt\hbox
   {$\rlap{\raise 5pt\hbox{$\char'074$}}\mathchar"7218$}}}   % < or of order
\def\simgreat{\mathbin{\lower 3pt\hbox
   {$\rlap{\raise 5pt\hbox{$\char'076$}}\mathchar"7218$}}}   % > or of order
\def\etal{{\rm et al.}}
\input psfig.sty
\def\solrad{{R$_\odot$}}

\def\solmas{{M$_\odot$}}
\def\solm{{M_\odot}}
\def\be{\begin{equation}}
\def\ee{\end{equation}}
\def\etal{et.al.}

\def \tcoll {t_{coll}}

\begin{document}

\title{The Formation of Massive Stars through stellar collisions}
\author{Ian A. Bonnell}
\affil{University of St Andrews, Physics and Astronomy, North Haugh, St Andrews, KY16 9SS, UK}

\begin{abstract}
In this review, I present the case for how massive stars may form
through stellar collisions. This mechanism requires very high stellar
densities, up to 4 orders of magnitude higher than are observed in the
cores of dense young clusters. In this model, the required stellar densities arise
due to gas accretion onto stars in the cluster core, including the
precursers of the massive stars. This forces the core to contract
until the stellar densities are sufficiently high for collisions to
occur. Gas accretion is also likely to play a major role in determining the
distribution of stellar masses in the cluster as well as the observed
mass segregation.  One of the main advantages of this mechanism is
that it explicitly relies on the cluster environment in order to
produce the massive stars. It is thus in a position to explain the
relation between clustered and massive star formation which is not an
obvious outcome of an isolated accretion mechanism. A recent numerical
simulation supports this model as the cluster core increases its
density by $10^5$ during gas accretion. Approximately 15 stellar
collisions occur (with $R_{coll}=1$ au) 
in the cluster core, making a significant
contribution to the mass of the most massive star.

\end{abstract}

\keywords{}

\section{Introduction} 
The formation of massive stars, due to their large chemical and
energetic feedback, is an important issue not just for the field of
star formation but also for many other areas of astronomy from the
interstellar medium to galaxy formation. The difficulties in forming
massive stars are two-fold. One, how does that much mass get
accumulated by one star, and two, can the accumulation of this matter
overcome the large radiation pressure from the star. The formation of
low-mass stars appears trivial in comparison as the stellar masses are
not vastly different from estimates of the Jeans mass in molecular
clouds and as the stellar luminosities are not sufficiently large to
affect the accretion process. Massive stars, in contrast, have masses
up to 100 times the estimated Jeans mass in molecular clouds. This
implies that the majority of their mass is probably due to subsequent
accumulation rather than the initial collapse (see A. Maeder, this
volume). Of potentially more of an impediment is the radiation
pressure from stars more massive then $\approx 10$ \solmas\ which can
act on the infalling dust grains to reverse the infall and halt the
accretion process (Yorke \& Krugel~1977; Yorke~1993; Beech \&
Mitalas~1994). There are possible solutions to this impediment
involving changing the properties of the dust grains (Wolfire \&
Cassinelli~1987) or perhaps a non-isotropic radiation flux such that
matter can more easily accrete onto the star's equator from an
accretion disc (c.f. H. Yorke, this volume). In this paper, I consideran alternative model whereby the accumulation of mass comes from stellar collisions.
The accumulation of such optically-thick, dust-free entities as stars elliminates any problem due to radiation pressure.

In constructing a theory for massive star formation, we should make
use of all the available observational clues. One such clue is that
massive stars are not generally found in isolation but appear to form
in the central regions of rich young stellar clusters (Zinnecker,
McCaughrean \& Wilking~1993; Hillenbrand~1997; Clarke, Bonnell \&
Hillenbrand~2000; Mermilliod~2001). Even runaway OB stars are probably
most easily explained as having been ejected from such a stellar
cluster (Clarke \& Pringle~1992). Testi \etal~(1997) have argued that
there is a strong correlation between the mass of the most massive
star in a cluster and the cluster's mean stellar density. Although
this data is still compatible with random pairing from an IMF (Bonnell \&
Clarke~1999), such a correlation would imply a potential causal
relationship between cluster density and the mass of the most massive
star. 

In considering the clustered environment for massive star formation,
there are a number of observations that need to be considered. In
addition to the large number of accompanying low-mass stars, the
relatively large stellar densities and that the mass spectrum is
similar to a field star IMF, one feature that needs to be explained is
the observed mass segregation (Hillenbrand~1997; Carpenter \etal~1997;
Hillenbrand \& Hartmann~1998). For example, the Orion Nebula cluster
(ONC) shows a segregation of massive stars towards the centre of the
cluster. This appears to be the case not only for the most massive O
stars but also for the intermediate-mass stars present in the cluster.

Lastly, but perhaps most importantly, the fact that massive stars form
in stellar clusters implies that we need to consider a dynamical
formation scenario. The dynamical time of the cluster as a whole is
comparable to the formation time of the individual stars. That implies
that the stars move around in the cluster while they are forming. We
have to thus abandon a static formation scenario where the environment
can be ignored. Although this causes significant complications, it can
also provide simple physical processes that combine to determine the
resultant stellar properties.

\section{Fragment masses}

The most straight forward solution for massive star formation would be
that a massive clump of $\approx 50$ \solmas\ collapses to form the
star. This scenario is complicated by the dense cluster
environment. As the massive stars are located in the cluster core, the
pre-stellar clump would have to fit in between the neighbouring stars,
which, in the core of the ONC are separated by $\approx 5000$ au. In
order for the clump to be gravitationally bound, it must have a Jeans
length smaller than this separation: \be R_J = \left(\frac{5 R_g T}{2
G \mu}\right)^{1/2} \left(\frac{4}{3} \pi \rho \right)^{-1/2} \simless
5000 {A.U.}, \ee where $\rho$ is the gas density and $T$ the
temperature, $R_g$ is the gas constant, $G$ is the gravitational
constant, and $\mu$ is the mean molecular weight.  This implies that
for typical temperatures in molecular clouds, the gas density must be
very high. This also makes sense as one would expect the highest gas
densities in the deepest part of the potential well (Zinnecker
\etal~1993).

The mass of the resultant star can then be estimated from the Jeans mass, the
minimum mass to be gravitationally bound:
\be
\label{Jeans_mass}
M_J = \left(\frac{5 R_g T}{2 G \mu}\right)^{3/2} \left(\frac{4}{3} \pi
\rho \right)^{-1/2}.  \ee In the core of the ONC, the Jeans mass is
then of order $0.3 \solm$, much smaller then the most massive star of
$50 \solm$, or the mean stellar mass of $5 \solm$ (Zinnecker
\etal~1993; Bonnell, Bate \& Zinnecker~1998). Something else is needed
in order to explain the observed masses.  Furthermore, from the above
equations it is obvious that the fragment mass should be smallest in
the centre of clusters as the gas density is likely to be largest
there. The initial masses from fragmentation in a stellar cluster
should therefore result in an inverse mass segregation, with small
stellar masses in the core (Bonnell, Bate \&
Zinnecker~1998).

Investigations of the dynamics of young stellar clusters have shown
that dynamical mass segregation, the sinking of massive stars due to
two-body relaxation, cannot explain the location of the massive stars
(Bonnell \& Davies~1998). Clusters such as the ONC are too young for
the massive stars that form the Trapezium to have sunk from a
significant distance from the cluster core. We are thus left with a
formation process that occurs in situ but that does not reflect the 
initial fragment or Jeans mass.

\section{Stellar collisions}

\begin{figure}[t]
%\vspace{-0.5truein}
\centerline{\psfig{figure=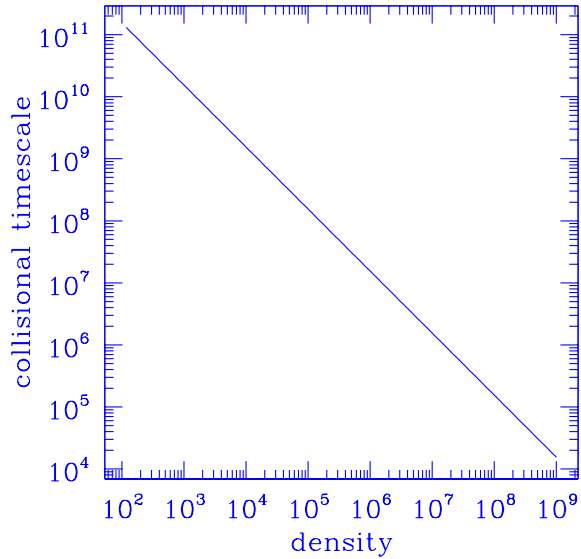,width=3.25truein,height=3.25truein}}
\caption{\label{fig1} The collisional timescale (in years) is plotted
as a function of the density in a stellar cluster in stars
pc$^{-3}$. A velocity dispersion of 2 km/s, a stellar mass of 10
\solmas, and a collisional radius of 10 \solrad\ are assumed. }
\end{figure}

If stellar collisions are responsible for the formation of massive
stars (eg. Bonnell, Bate \& Zinnecker~1998), then the timescale for multiple collisions has to be less than
the observed age of the massive stars. Although it is difficult to
ascertain the exact ages of massive stars, we know that in Orion, the
massive stars are younger than a few million years and potentially as
young as a few $\times 10^5$ years. The collisional timescale $\tcoll$ is related to the stellar properties as (Binney \& Tremaine~1987)
\begin{equation}
{1 \over \tcoll} = 16 \sqrt{\pi} v_{\rm disp}R_*^2 \large(1+ {GM_*\over 2 
 v_{\rm disp}^2 R_*}\large),
\end{equation}
where $n$ is the density of stars in the cluster, $v_{\rm disp}$ is
the velocity dispersion, and $M_*, R_*$ are the mass and radius of the
star undergoing the collision. Figure~\ref{fig1} plots the collisional
timescale as a function of stellar density for collisions involving a
10 \solmas\ star assuming a 10 \solrad\ collisional radius. The
velocity dispersion is assumed to be 2 km/s. We can see from Figure~\ref{fig1}
that for densities typical of young stellar clusters ($n \approx 10^3$
stars pc$^{-3}$), the collisional timescale is $> 10^{10}$ years. In
order to get a collisional timescale of order $\simless 10^{5}$ years,
a stellar density of $\simgreat 10^8$ stars pc$^{-3}$ is required.

Young stellar clusters are relatively dense agglomerations of
stars. The Orion Nebula Cluster (ONC), for example, has a mean stellar
density of $n\approx 10^3$ stars pc$^{-3}$ (Hillenbrand~1997). The
core of the ONC is considerably more dense with $n\approx 2 \times
10^4$ stars pc$^{-3}$ (McCaughrean \& Stauffer~1994). From Figure~1 we
see that the corresponding collisional timescale is $\approx 7 \times
10^8$ years.  Clearly collisions cannot play an important role in the
present conditions of the ONC. If stellar mergers are an important
process in massive star formation, then the required stellar density
of $n \simgreat 10^8$ stars pc$^{-3}$ implies that the ONC would have
had to undergo a high density phase. The main question is then what
process could have lead to such a short-lived but high-density
phase. In the following sections we consider how accretion in stellar
clusters may lead to such a scenario.

\section{Accretion and the IMF}

\begin{figure}[t]
%\vspace{-0.5truein}
\centerline{\psfig{figure=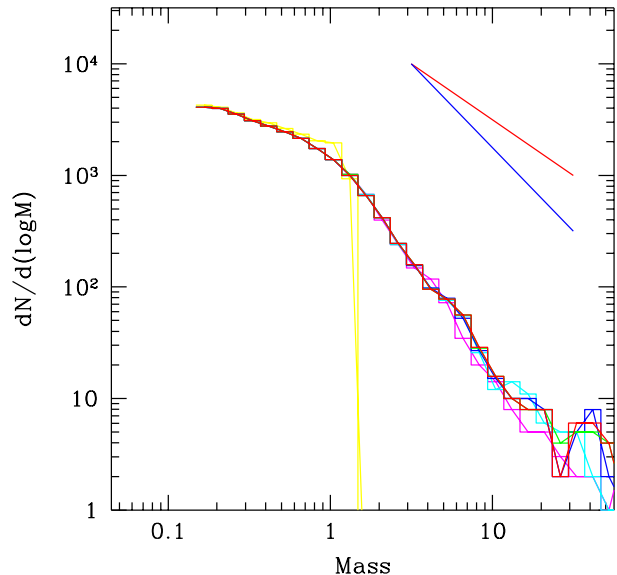,width=3.75truein,height=3.75truein}}
\caption{\label{crittime} The Initial mass function is plotted that results from an accreting
cluster. The accretion is assumed to be  tidal accretion when the gas
dominates the cluster potential 
followed by Bondi-Hoyle accretion when the stars dominate the potential in the
 cluster core. The accretion results in a $\gamma=-1.5$ IMF for low-mass stars and a steeper $\gamma\approx -2.5$ for high-mass stars (Bonnell \etal~2001b).}
\end{figure}

Observations of young stellar clusters reveal that they generally
contain significant amounts of mass in the form of gas. This gas can
amount up to 90 \% of the total cluster mass (Lada~1991). Accretion of
this gas can thus be an important factor in determining the masses of
individual stars as well as the properties of the stellar cluster.

Numerical simulations of accretion in clusters have found that
accretion naturally results in a mass segregated cluster (Bonnell
\etal~1997,~2001a). This occurs as the cluster potential funnels the
gas down to the centre of the cluster where it is accreted by whatever
stars happen to be there. The non-uniform accretion rates then result
in the most massive stars being located in the centre of the cluster.
In general, the resulting cluster demonstrates a degree of mass segregation 
comparable to that found in young stellar clusters.

Another finding of these numerical simulations is that there are two
different physical regimes for the accretion, depending on whether the
gas or the stars dominate the gravitational potential (Bonnell
\etal~2001a). When the gas dominates the potential, the motions of the
stars and the gas are determined by the changing gravitational
potential as the gas collapses. In this regime, the stellar and gas
motions are similar and the accretion rate is determined by each
star's tidal radius relative to the cluster potential. Material that
moves inside the tidal radius gets accreted by the star.  This assumes
that the gas is initially unsupported, as is expected if it is able to
fragment to form the large number of stars in the cluster.

The second regime occurs when the stars dominate the gravitational
potential.  In this case, the stars virialise and have motions
uncorrelated to those of the gas. The gas velocity relative to a star
is therefore large such that it is the determining factor as to
whether the gas is bound to the star. The accretion rate in this case
is the common Bondi-Hoyle accretion which depends on the local gas
velocity and the star's mass, but not on the external potential.

Using these two different regimes for the gas accretion, Bonnell
\etal~(2001b) showed how accretion in stellar clusters can result in a
double power-law IMF where the low-mass stars have a shallow
$\gamma\approx -1.5$ (where Salpeter is $\gamma=-2.35$) power-law due to the
tidal lobe accretion while the high-mass stars have a steeper
power-law ($\gamma\approx -2.5$) due to the subsequent Bondi-Hoyle accretion.
The high-mass stars accrete the majority of their mass in the stellar dominated regime as they form in the core of the cluster where the stars (due to their
high accretion rates) first come to dominate the potential.

\section{Accretion and Cluster Dynamics}

\begin{figure}[t]
%\vspace{-0.5truein}
\centerline{\psfig{figure=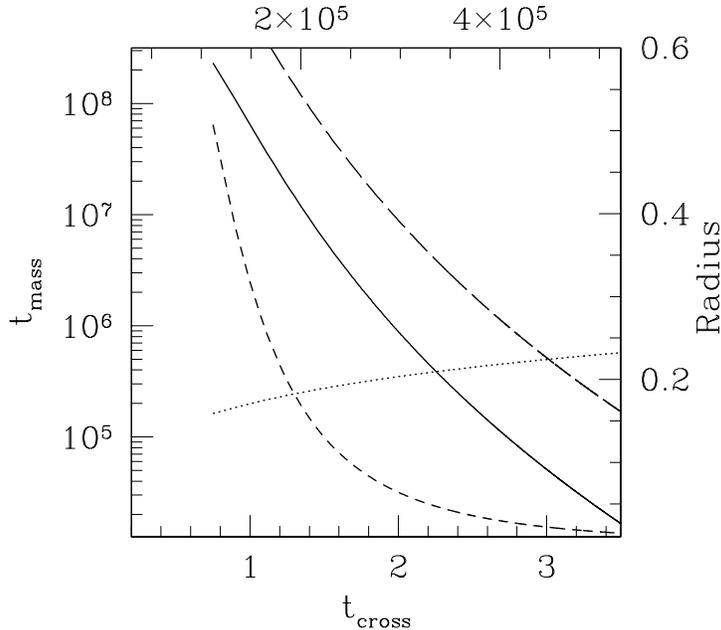,width=4.0truein,height=4.0truein,rwidth=3.75truein,rheight=3.75truein}}
\caption{\label{crittime} The evolution of an accreting cluster is
shown where the cluster shrinkage is modelled by equation~(6). The
accretion timescale (dotted line) increases while the collisional
timescales (solid and long-dashed lines) decrease reflecting the
change in cluster radius (short-dashed line).The collisional radii are
taken to be $0.1$ and $0.01$ au. (from Bonnell \etal~1998)}
\end{figure}

The simulations in the previous section also found that accreting clusters tend to shrink as mass is added onto the stars (Bonnell \etal~2001a). In order
to understand why this happens let us consider the energy of the stars in the
cluster potential (c.f. Bonnell, Bate \ Zinnecker~1998):
\begin{equation}
E_{stars}= {p^2\over 2M_{stars}} - {GM_{stars}M_{tot} \over R},
\end{equation}
where $p$ is the total momentum of the stars, $M_{stars}$ is the mass in
stars, $M_{tot}$ is the total mass in both stars and gas while $R$ is
the size of the system.  In what follows we assume that the accreting
gas has no net momentum and that therefore the stellar momenta is
conserved during accretion.  This assumption is justified as in
general the accreting gas will be spherically infalling and, once the
stars are virialised, there is no correlation between gas motions
and stellar motions.

As mass is added while momentum is conserved, the kinetic energy of the
stars decreases at the same time as the stars become more bound to the
potential. The stars are then no longer in virial equilibrium with the 
potential and thus shrink until they revirialise at a smaller radius.
The new radius reflects the new, lower energy as:
\begin{equation}
R_{cluster} = -{GM_{tot}M_{stars} \over 2 E_{new}}.
\end{equation}

As long as the accretion timescale is longer than the dynamical
timescale, then the cluster can accrete and revirialise many times and
thus shrink substantially. If, on the other hand, the accretion timescale is shorter than the dynamical timescale, then the cluster can only shrink by a factor of 2 after the accretion has finished.
Under the condition of slow accretion, the rate of cluster shrinkage can be
estimated by considering that accretion occurs at a constant radius and then
shrinks to revirialise, reflecting the new energy. Thus
\begin{equation}
\left({\partial E \over \partial M_{stars}}\right)_R = \left({d E_{new} \over d M_{stars}}\right). 
\end{equation}

Solving the above equation yields (Bonnell \etal~1998)
\begin{equation}
R \propto M_{stars}^{-\alpha},
\end{equation}
where the value of $\alpha$ depends on whether the total mass is conserved
($\alpha =2$ or whether it is proportional to the mass in stars ($\alpha=3$).
The latter possibility reflects the case when accretion occurs onto the core
of the cluster and gas continues to infall from outside the core.
Thus, the increase in the stellar density can be very large, given by
\begin{equation}
n \propto M_{stars}^{3 \alpha},
\end{equation}
such that for $\alpha=3$ and an increase in the stellar mass by a factor
of three results in an increase in the density as $3^9\approx 2\times 10^4$.

We therefore take as our model the core of a cluster similar to the
ONC.  This core, containing $\approx 100$ initially low-mass stars
within $\approx 0.1$ pc. The stars in the core accrete until they have
reached intermediate masses ($\approx 5 \solm$) by which time they
have attained a sufficiently high density that stellar collisions and
mergers occur to form the massive stars. Figure~\ref{crittime} shows the evolution
of such a system as the stars accrete at a constant accretion rate of
$2 \times 10^{-6} \solm$ year$^{-1}$(from Bonnell \etal~1998). 
The half-mass radius of the
system is plotted as a function of time as are the collisional
timescale and the accretion timescale. Once the cluster core has
shrunk sufficiently such that the collisional timescale is shorter
than the accretion timescale, stellar mergers dominate the mass
buildup process. Two collisional timescales are plotted reflecting
collisional radii of $0.1$ and $0.01$ au. The transition from
accretion to mergers occurs after a few crossing times, corresponding
roughly to $4 \times 10^5$ years.

\section{Stellar Collisions and Binaries}

The actual cross sections for collisions depends on the dynamics and
the presence of circumstellar material. One of the crucial factors is
the formation of binaries as they then have a much larger cross
section for further interactions. There are two ways in which binary
systems can form. Firstly, they can form through tidal capture
encounters.  This is most important for near-equal mass
encounters. Most encounters involving the more massive stars in the
core are with lower-mass stars.  These encounters do not generally
result in tidal captures due to the disparate interior densities of
the stars. Instead, the lower-mass star can be tidally disrupted to
form a circumstellar disc. This results in the second binary formation
mechanism as a subsequent encounter that passes through the disc gets
captured due to the binding energy of the disc (Davies \etal, in
preparation).  Such binary systems will either merge on subsequent
encounters or exchange in a higher-mass star to form a near equal-mass
binary, that may itself merge in further encounters.

In either case, a high frequency of near-equal mass binary systems should
result from a system where stellar collisions are occurring. This agrees
with the finding that many O stars are members of short-period binary systems
(Mason \etal~1996). This is also the case in O-stars found in the centre of open clusters where they all appear to be in close binary systems with stars of comparable masses (Mermilliod~2001).

\section{Accretion and Cluster Dynamics: the UKAFF simulation}

\begin{figure}[t]
%\vspace{-0.5truein}
\centerline{\psfig{figure=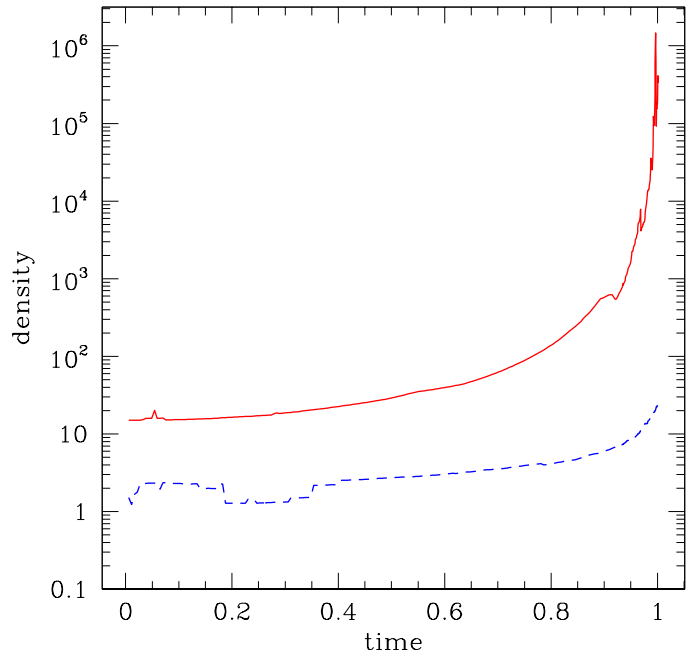,width=4.0truein,height=4.0truein,rwidth=3.75truein,rheight=3.75truein}}
\caption{\label{clusdenevol} The evolution of the mean (dashed line)
and maximum (solid line) stellar density is plotted as a function of
time in units of the initial free-fall time.  The maximum density
increases dramatically near the end of the simulation, reaching values
at which stellar collisions can occur in significant numbers. The
maximum density is calculated by computing the volume needed to
contain 10 stars while the mean density is averaged over stars near
the half-mass radius.}
\end{figure}
\begin{figure}[t]
%\vspace{-0.5truein}
\centerline{\psfig{figure=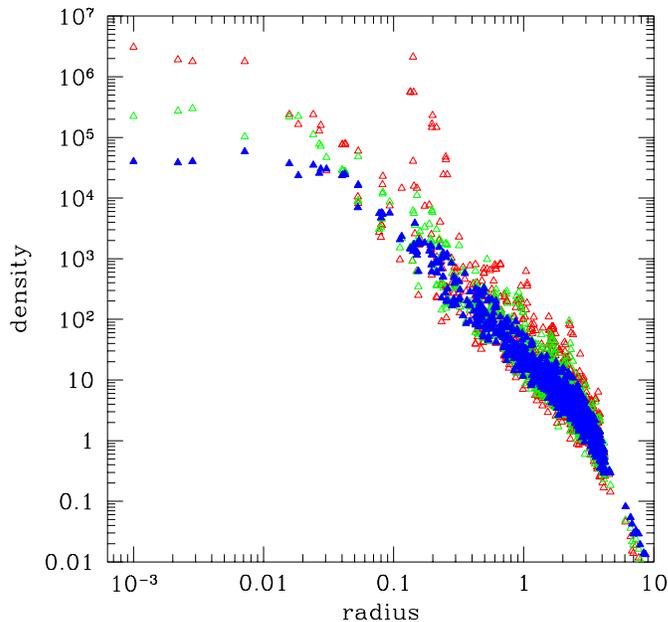,width=3.75truein,height=3.75truein}}
\caption{\label{stellrad} The distribution of stellar densities at the end of the UKAFF simulation is plotted as a function of radius in the cluster.  The density is calculated by computing the volume needed to contain (from the top) 3, 5 (both open triangles) or 10 stars (filled triangles).
The half-mass radius of the cluster is at $r\approx 1$.}
\end{figure}

\begin{figure}[t]
%\vspace{-0.5truein}
\centerline{\psfig{figure=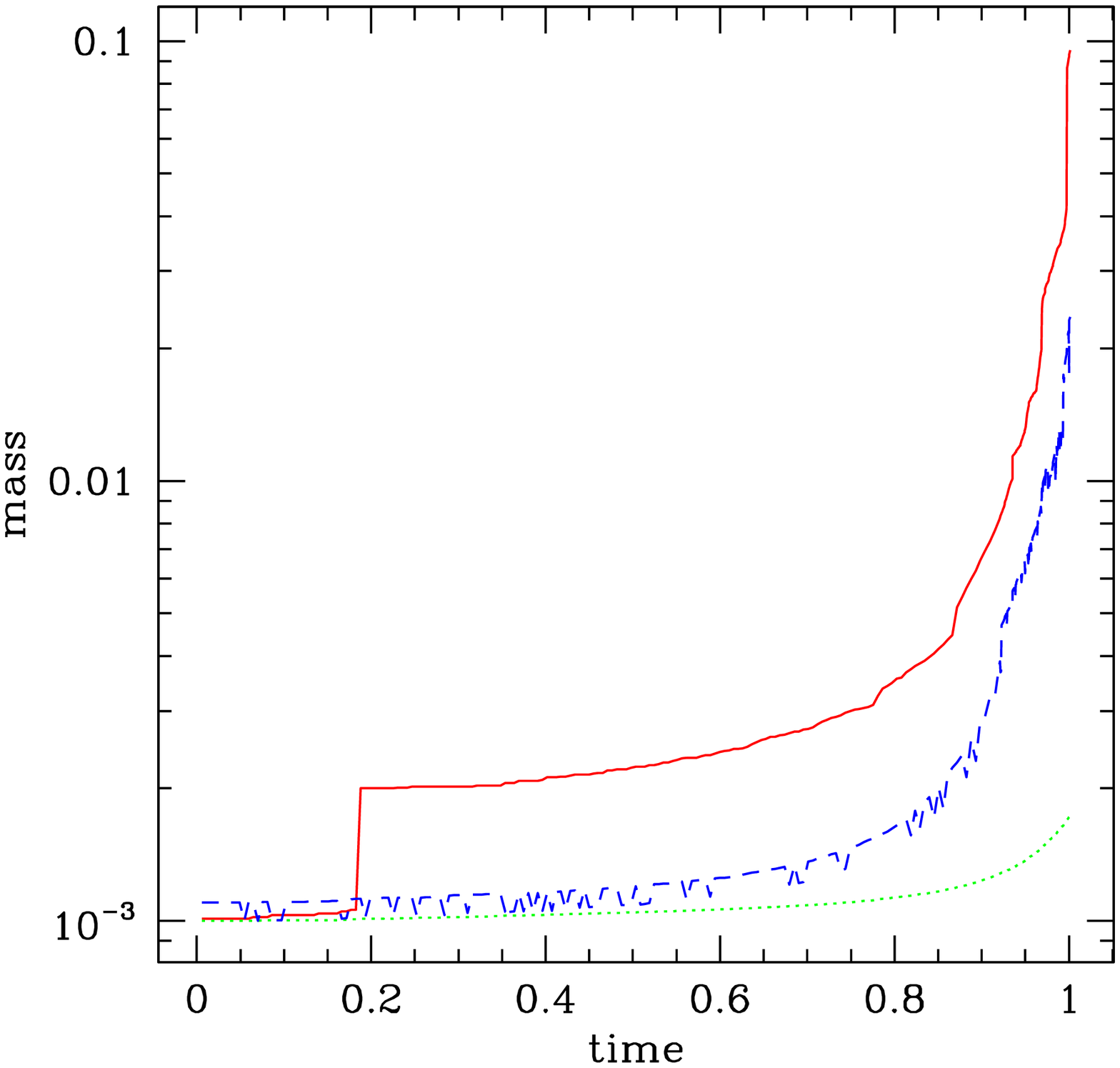,width=4.0truein,height=4.0truein,rwidth=3.75truein,rheight=3.75truein}}
\caption{\label{maxmassevol} The evolution of the mean (dotted line)
and maximum (solid line) stellar mass is plotted as a function of
time in units of the initial free-fall time.  The mean stellar mass inside the cluster core is also plotted (dashed line). The maximum mass increases
in jumps every time there is a merger and smoothly due to accretion. Note that there are two major mergers near the end of the simulation which contribute
significantly to the maximum mass. The maximum mass increases by a factor of $~100$ while the mean stellar mass has increased by only 70 \%. The mean stellar
mass in the core also increases dramatically signifying a mass segregated cluster.}
\end{figure}

In order to study the dynamics and evolution of an accreting cluster
we have performed a numerical simulation of gas accretion in a cluster
containing 1000 stars (Bonnell \& Bate, in preparation). This simulation,
performed on the UKAFF supercomputer with $10^6$ SPH particles,
follows the gas and stellar dynamics as the stars accrete. The primary
aims of this simulation were to investigate whether an actual stellar
and gas dynamical system would react according to our simple
prescription and to see if the stellar interactions would destroy the
cluster core before the necessary densities for collisions occurs.
The maximum density in the cluster, as well as the density at the
half-mass radius are plotted as a function of time in
Figure~\ref{clusdenevol}. We see that the density increases
dramatically after about one free-fall time. The mean density of the
cluster increases by a factor of $\approx 20$ while the peak density
increases by over $10^4$ up to $10^5$. This large increase in stellar
density implies a large decrease in the timescale for collisions and
mergers. If we take the cluster's initial conditions as being those
present in the ONC, then we have an initial mean density of $10^3$
stars pc$^{-3}$ and a peak density of $2 \times 10^{4}$ stars
pc$^{-3}$ (McCaughrean \& Stauffer~1994). In the UKAFF simulation the
peak density increases by a factor of $>10^5$ times the initial mean
density. This then corresponds to stellar densities of $n \simgreat
10^8$ stars pc$^{-3}$. This stellar density is sufficient to produce
significant numbers of stellar collisions in the cluster core.

Accretion results in a centrally condensed cluster with a $n\propto r^{-2}$
stellar density profile. Figure~\ref{stellrad} plots the distribution of stellar densities calculated as the volume necessary to contain 10 stars. The
maximum density is considerably higher than the mean density of $n\approx 20$,
but is contained in a core which occupies only the innermost $\approx 0.01$ of
the half-mass radius.

In the UKAFF simulation, collisions are assumed to occur when two
stars pass within 1 au of each other. This value is much larger than
the stellar radii but comparable to the binary separations considered
above. In any case, as shown in Bonnell \etal~(1998), the actual
collisional radius affects only the delay until the cluster has
shrunk sufficiently for collisions. Using the above collisional
radius, 15 collisions occur during the UKAFF simulation. About half of
these involve the more massive stars including a near equal-mass
collision which results in the most massive star having a mass of $\approx 100$
times the initial stellar masses. We thus can conclude that an
accreting cluster can contract far enough such that significant
stellar collisions occur which result in the formation of massive
stars.  The evolution of the mean and maximum masses during the
simulation are plotted in Figure~\ref{maxmassevol}. We see that the
maximum mass increases dramatically through both accretion (smooth
increase) and collisions (jumps).  As noted above, the maximum mass
reaches a value $\approx 100$ times the initial stellar mass. At the
same time the mean stellar mass only increases by $\approx 70$\%
during the evolution. Many of the mergers involving the most massive
star involve lower-mass impactors as is expected. Two mergers near the
end of the simulation involve near-equal mass components and thus
significantly contribute to the mass.  The mean stellar mass in the cluster core
is also plotted in Figure~\ref{maxmassevol}. This mean mass in the core
increases dramatically but is only half due to the most massive star.
Thus the cluster has been mass segregated due to the accretion and collisions.

It is also worth noting that the cluster stars tend to be in small
groups, and along filaments (see figure~\ref{gasstreams} below). These
configurations increase collisional rates and thus the ability
for collisions to form massive stars. Furthermore, if the absence of
collisions, the maximum stellar densities would have been considerably
higher.

\section{Collimating Outflows}

\begin{figure}[t]
%\vspace{-0.5truein}
\centerline{\psfig{figure=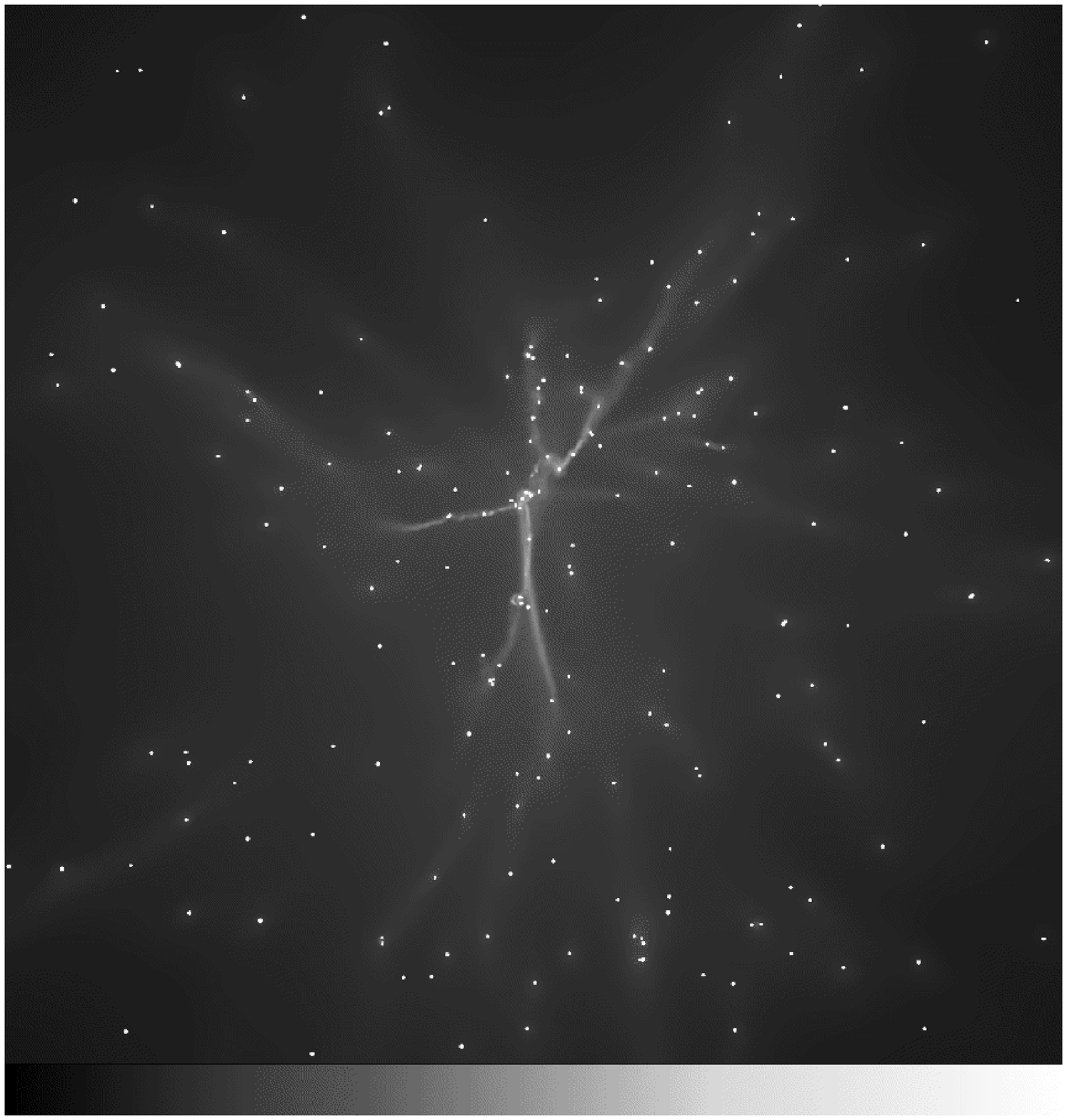,width=4.75truein,height=4.75truein}}
\caption{\label{gasstreams} The distribution of gas and stars are
shown near the end of the UKAFF simulation. The region plotted is the
central core of radius one half the half-mass radius. The gas forms
into filaments which may help collimate any outflows.}
\end{figure}

Massive stars are generally associated with energetic outflows. These
outflows are often seen as evidence for a circumstellar disc which
would most likely be destroyed in regions sufficiently dense to allow
for stellar mergers.  The implication of a stellar disc for these
outflows is not at all certain as the outflows are generally poorly
collimated and thus of a fundamentally different type than the
well-collimated outflows from low-mass stars. 

An alternative to disc collimation is that there is present some
structure in the accretion flow or general circumstellar environment
which tends to collimate (poorly) any outflow from the massive
stars. Figure\ref{gasstreams} shows the configuration of the stars and
the gas towards the end of the UKAFF run. The gas (and stars) are far
from uniformly distributed. Instead, the gas displays filamentary
shapes pointing towards the cluster centre. The stars are generally
situated along the filaments and tend to be in small groups. Both the
gas and the stars move along these filaments which increases the
rate of collisions.

The filamentary structures present in the accretion flows in the
cluster centre are due to gravitational instabilities. The gas
initially contains some 1000 Jeans mass, in agreement with its
previous fragmentation to form the 1000 stars. In such a system,
gravity tends to increase any over densities and to reduce the
dimensions of the structure. In this way, the over densities can grow
and form into filamentary structures. Such structures are a possible
source for the collimation of the massive stars' outflows. Further
numerical simulations will have to be performed in order to test this
speculation. It should be noted, however, that any collimation that
does result from this structure is likely to be fairly open and as
such this mechanism is not likely to result in the formation of
jet-like outflows. Any such highly collimated outflows around young O
stars would thus be difficult to explain with this stellar merger
mechanism.

\section{Implications of collisions}

There are many implications of this collisional formation mechanism
for massive stars which need to be considered. The most significant of
these is the effect of the energy involved in the collisions. As two
$10 \solm$ stars collide, the energy involved is of order $10^{49}$
ergs. This much energy will puff up the stars which will contract on
their Kelvin-Helmhotz timescale. The increased size of the star will
make it more susceptible to further collisions.  if we assume that the
energy is radiated over some $10^4$ years, then the energy flux will
be $\approx 10^{38}$ ergs s$^{-1}$. Indeed, a fraction of the kinetic
energy may be released on much shorter timescales and result in an
x-ray flare. Such an event should be observable (see chapters by
H. Zinnecker and J. Bally). It is alkso probable that the energy
released in this way would contribute to the re-expansion of the
cluster core.

\section{Summary}

The formation of massive stars is a complicated process in which the
environment is likely to play a crucial role. Massive stars form in the cores
of stellar clusters where the expected fragment mass is low. Accretion in clusters  forms higher-mass stars from low mass stars and naturally results in a mass segregated cluster as the higher accretion rates in the core result in more massive stars there. This process also results in a two-slope IMF similar to that observed in stellar clusters and in the field. 

The accretion onto the stars in the cluster forces it to shrink and
revirialise at a smaller radius. if the accretion is slower or
comparable to the local dynamical time (say of the cluster core), then
accretion can force the core to continue shrinking until it has
reached sufficient densities where stellar collisions occur. These
collisions can then result in the formation of massive stars.  A
recent simulation using the UKAFF supercomputer has shown that an
accreting cluster can indeed attain the required stellar densities and
that mergers can play an important role in the formation of massive
stars.  The accretion flow is susceptible to the formation of
filamentary structure which may play a crucial role in collimating any
outflow from massive stars.  It would be unlikely to be able to
collimate a jet-like outflow.

\section{Acknowledgments}
The results presented here are partially based on numerical simulations 
performed with the 
United Kigndom's Astrophysical Fluid Facility (UKAFF) supercomputer.

\end{document}